\documentclass[letterpaper]{article} 
\usepackage{aaai2026}  
\usepackage{times}  
\usepackage{helvet}  
\usepackage{courier}  
\usepackage[hyphens]{url}  
\usepackage{graphicx} 
\usepackage{natbib}  
\usepackage{caption} 
\usepackage{algorithm}
\usepackage{algorithmic}
\usepackage{amsmath}
\usepackage{amsfonts}
\usepackage{booktabs}
\usepackage{multirow}
\usepackage{array}
\usepackage{times}
\usepackage{helvet}
\usepackage{courier}
\usepackage{xcolor}
\usepackage{svg}
\usepackage{newfloat}
\usepackage{listings}
\DeclareCaptionStyle{ruled}{labelfont=normalfont,labelsep=colon,strut=off} 
\floatstyle{ruled}
\newfloat{listing}{tb}{lst}{}
\floatname{listing}{Listing}
\title{Cross-Space Synergy: A Unified Framework for Multimodal Emotion Recognition in Conversation}
\author{
    Xiaosen Lyu\textsuperscript{\rm 1,2}\equalcontrib,
    Jiayu Xiong\textsuperscript{\rm 1,2}\equalcontrib, 
    Yuren Chen\textsuperscript{\rm 1,2}, 
    Wanlong Wang\textsuperscript{\rm 1,2}, 
    Xiaoqing Dai\textsuperscript{\rm 1,2}, 
    Jing Wang\textsuperscript{\rm 1,2}\thanks{Corresponding author.}\\
}
\affiliations{
    \textsuperscript{\rm 1}School of Computer Science and Technology, Huaqiao University, Xiamen, China\\
    \textsuperscript{\rm 2}Xiamen Key Laboratory of Computer Vision and Pattern Recognition, Xiamen, China

    xslyu@stu.hqu.edu.cn, wroaring@hqu.edu.cn
}

\usepackage{bibentry}

\begin{document}

\maketitle

\begin{abstract}
Multimodal Emotion Recognition in Conversation (MERC) aims to predict speakers' emotions by integrating textual, acoustic, and visual cues. Existing approaches either struggle to capture complex cross‑modal interactions or experience gradient conflicts and unstable training when using deeper architectures. To address these issues, we propose \textbf{Cross-Space Synergy (CSS)}, which couples a representation component with an optimization component. \textit{Synergistic Polynomial Fusion (SPF)} serves the representation role, leveraging low-rank tensor factorization to efficiently capture high-order cross-modal interactions. \textit{Pareto Gradient Modulator (PGM)} serves the optimization role, steering updates along Pareto-optimal directions across competing objectives to alleviate gradient conflicts and improve stability. Experiments show that CSS outperforms existing representative methods on IEMOCAP and MELD in both accuracy and training stability, demonstrating its effectiveness in complex multimodal scenarios.
\end{abstract}


\section{Introduction}
With the increasing deployment of affective computing in human-computer interaction, emotion recognition has garnered significant attention. Early research primarily addressed unimodal Emotion Recognition in Conversation (ERC) through textual cues~\cite{alm-etal-2005-emotions}, overlooking behavioral signals from speech and visual cues. The emergence of multimodal conversational data has redirected attention toward the more challenging Multimodal Emotion Recognition in Conversation (MERC) task~\cite{chen2017multimodal}, which demands integrating heterogeneous text, audio, and visual signals to accurately infer speakers' emotional states. Compared to ERC, MERC exhibits greater complexity due to multimodal heterogeneity and multiple concurrent training objectives, including primary classification and auxiliary supervision. These complexities necessitate unified modeling strategies that simultaneously enhance cross-modal representation learning and stabilize multi-objective optimization.

Prior MERC research has explored modality-specific representation learning and cross-modal interaction modeling, yielding substantial progress. Attention-based models~\cite{Majumder2018DialogueRNNAA, nguyen2024ada2i, ma2024sdt} integrate multimodal information using self-attention or cross-attention, but often rely on shallow fusion strategies that restrict their ability to capture high-order semantic dependencies. Multi-stage fusion frameworks~\cite{liu-etal-2018-efficient-low, tsai2019multimodal} enhance modular interpretability through staged integration, but sacrifice holistic multimodal alignment. Despite their differences, these approaches share a common limitation: their fusion mechanisms are too shallow to capture the nonlinear, high-order interactions critical for modeling complex emotional dynamics in conversation.

Existing studies have sought to overcome shallow fusion by enhancing model expressiveness via deeper architectures or more structured interaction mechanisms. Graph-based methods~\cite{ghosal-etal-2019-dialoguegcn, 10203083}, for instance, enhance cross-modal reasoning by constructing speaker or modality graphs and propagating messages through multiple hops. However, these designs often rely on predefined graph topology and rigid structural assumptions, leading to limited flexibility and high inference overhead. Moreover, deeper or more expressive architectures frequently encounter optimization instability~\cite{he2016deep, sun2022surprising}, especially in settings with multiple learning objectives.

In the context of MERC, this manifests as semantic entanglement across tasks and conflicting gradient signals, where objectives such as multimodal classification, unimodal consistency, and auxiliary guidance compete for shared features~\cite{sener2018multi}. As a result, increasing network capacity may not improve performance, but instead hinder convergence and generalization. This creates a fundamental dilemma: shallow interaction leads to under-expressive modeling, while deeper fusion introduces chaotic optimization dynamics—posing a trade-off between representational expressiveness and training stability.

To systematically address this dilemma, we reconceptualize MERC as a cross-space synergy problem from two interconnected dimensions. In representation space, deeper fusion structures provide enhanced nonlinear modeling but introduce semantic entanglement—different objectives compete for shared feature spaces, creating conflicts between specificity and generality. In gradient space, multiple learning objectives frequently conflict, where traditional joint optimization strategies may disrupt primary task learning trajectories. Therefore, MERC requires transcending traditional paradigms, necessitating a unified framework that coordinates representation construction and gradient alignment to ensure mutual reinforcement during training. Based on this insight, we propose the \textbf{Cross-Space Synergy (CSS)} framework to coordinate conflicting issues across different spaces. Our main contributions are as follows:

\begin{itemize}
    \item \textbf{We propose Synergistic Polynomial Fusion (SPF) for expressive representation-space fusion}, which models nonlinear cross-modal interactions via structured tensor composition, enhanced by modality-specific projection and stability-oriented mechanisms.

    \item \textbf{We design the Pareto Gradient Modulator (PGM) for stable gradient-space optimization}, dynamically selecting Pareto-optimal gradients to balance conflicts among multimodal classification, unimodal regularization, and distillation objectives.

    \item \textbf{Extensive experiments on IEMOCAP and MELD demonstrate} that our CSS framework consistently outperforms strong baselines, achieving enhanced accuracy and training stability even with lower fusion complexity.
\end{itemize}

\section{Related Work}
\subsection{Multimodal Emotion Recognition in Conversation}
Early emotion recognition in conversation (ERC) studies primarily relied on textual cues, employing recurrent and memory-based architectures to capture speaker dynamics and contextual dependencies~\cite{Majumder2018DialogueRNNAA,8764449}.The emergence of audio-visual corpora has driven research toward Multimodal ERC (MERC), where models integrate heterogeneous text, audio, and visual signals for enhanced emotional understanding. MERC introduces significant challenges, including cross-modal heterogeneity, temporal dynamics, and multi-objective supervision. Recent advances have explored diverse architectural strategies: DialogueGCN employs graph-based modeling for speaker interactions~\cite{ghosal-etal-2019-dialoguegcn}, MulT utilizes cross-modal attention ~\cite{tsai2019multimodal}, MISA incorporates modality-invariant and specific representations~\cite{Hazarika2020MISAMA}, while SDT introduces auxiliary branches with self-distillation~\cite{ma2024sdt}. Despite performance gains, existing methods remain limited in fusion expressiveness and training stability due to uncoordinated multimodal objectives. Critical gaps remain in dynamic modality coordination, context-aware adaptation, and robust cross-modal representation learning—core requirements for scalable multimodal conversational systems.

\subsection{Multimodal Fusion Methods}
Multimodal fusion constitutes a core component in MERC, aiming to integrate heterogeneous text, audio, and visual signals into unified representations. Early fusion strategies employed feature concatenation, which fails to capture semantic interactions between modalities. Advanced approaches have incorporated gating mechanisms and attention-based fusion to enhance cross-modal coordination. Gated fusion methods modulate information flow through modality-specific weighting~\cite{Zadeh_Liang_Mazumder_Poria_Cambria_Morency_2018}, while cross-modal attention facilitates explicit semantic alignment~\cite{tsai2019multimodal}.

Recent work has explored bilinear models and high-order tensor interactions to address expressiveness limitations. Methods such as MLB~\cite{fukui2016mlb}, MFB~\cite{yu2017mfb}, and LMF~\cite{liu-etal-2018-efficient-low} employ multiplicative fusion operations (e.g., Hadamard products, low-rank factorization) to model nonlinear cross-modal dependencies, though these remain constrained to pairwise interactions. High-order approaches like Polynomial Tensor Pooling (PTP)~\cite{NEURIPS2019_f56d8183} leverage tensor decomposition to compactly represent multi-way modality interactions. However, existing methods predominantly utilize shared projection parameters, limiting their capacity to capture modality-specific structural properties and lacking explicit mechanisms for controlling contribution strength, resulting in sensitivity to noisy or imbalanced multimodal inputs.

\subsection{Multi-Objective Optimization Strategies}
Multimodal emotion recognition tasks inherently involve multiple learning objectives with distinct representational foci and gradient dynamics. Naive aggregation through static weighted losses frequently induces gradient interference, resulting in unstable convergence or single-task dominance. Coordinating these conflicting optimization signals has emerged as a fundamental challenge in multi-task learning paradigms. Traditional approaches depend on manual hyperparameter tuning or predetermined weight scheduling, which prove inadequate under dynamic and imbalanced training regimes.

Recent advances have investigated multi-objective optimization (MOO) frameworks that pursue gradient-level equilibrium solutions. The Multiple Gradient Descent Algorithm (MGDA) \cite{desideri2012multiple,sener2018multi} exemplifies this direction by computing Pareto-optimal task gradient fronts and identifying shared descent directions that benefit all objectives simultaneously. Subsequent extensions, including PCGrad\cite{yu2020pcgrad} and CAGrad~\cite{NEURIPS2021_9d27fdf2}, mitigate gradient conflicts through projection-based and constraint-driven strategies, respectively. Despite demonstrated efficacy in general multi-task scenarios, existing MOO methods are predominantly designed for unimodal or homogeneous task settings and operate independently of fusion mechanisms—thereby limiting synergistic potential in multimodal architectures such as MERC systems.

\section{Methodology}
\begin{figure*}[t]
    \centering
    \includegraphics[width=0.94\textwidth]{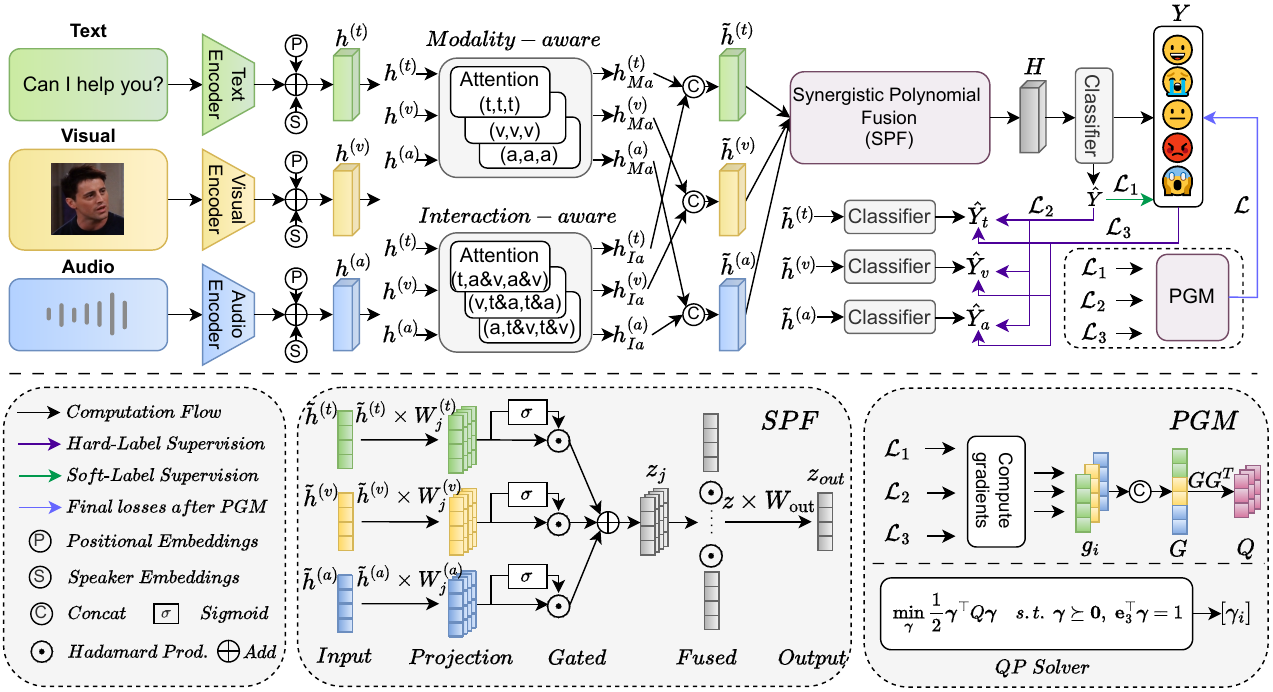} %
    \caption{
    Overview of the proposed CSS framework, which integrates two-stage representation encoding, high-order multimodal fusion (SPF), and multi-objective optimization (PGM).
    }
    \label{fig2}
    \end{figure*}

    \subsection{Overview}

    We address the task of MERC, which aims to predict the emotional label of each utterance in a dialogue by leveraging \textbf{text}, \textbf{audio}, and \textbf{visual} features under conversational context. Formally, given a dialogue consisting of $T$ utterances $\mathcal{C} = \{u_1, u_2, ..., u_T\}$, each utterance $u_t$ is associated with modality-specific features $\mathbf{x}_t^{(m)}$, where $m \in \{\text{text}, \text{audio}, \text{visual}\}$. The objective is to learn a function $f$ that maps each utterance and its context to a label space $\mathcal{Y}$.
    \begin{equation}
    \hat{y}_t = f(\mathcal{C}, \{\mathbf{x}_t^{(m)}\}), \quad \hat{y}_t \in \mathcal{Y}.
    \end{equation}
    
    MERC is challenging due to dynamic dialogue structure, speaker interactions, and the need for multimodal synergy. Moreover, training often involves multiple objectives whose gradients conflict, causing unstable or slow convergence. Prior work typically simplifies network structure to mitigate this, but at the cost of limited modeling capacity.
    
    To overcome these issues, we propose \textbf{Cross-Space Synergy (CSS)}, a unified framework that enhances both representation and optimization. As shown in Figure~\ref{fig2}, CSS comprises three components: \textbf{(1) Representation Encoding}, a speaker- and context-aware encoder; \textbf{(2) Representation-Space Fusion}, where \textit{Synergistic Polynomial Fusion (SPF)} models high-order cross-modal interactions via low-rank tensor composition; and \textbf{(3) Gradient-Space Optimization}, where \textit{Pareto Gradient Modulator (PGM)} dynamically coordinates multiple learning signals through Pareto-optimal descent. This synergy enables expressive yet stable multimodal emotion modeling.

    \subsection{Representation Encoding}
     To capture both intra-modal structure and cross-modal interactions, we design a compact two-stage encoder consisting of \textit{modality-aware} and \textit{interaction-aware} attention modules. Inspired by recent attention-based methods~\cite{10502283,ma2024sdt}, we adopt similar mechanisms but organize them into a tailored two-stage structure.

     Before encoding, each input is enriched with a learnable speaker embedding~\cite{hu-etal-2021-mmgcn} and sinusoidal positional encoding~\cite{vaswani2017attention}. These are projected into a shared $d$-dimensional space via modality-specific 1D convolutions, yielding initial utterance representations $\mathbf{h}^{(m)} \in \mathbb{R}^d$.
     
     \textbf{Modality-aware Encoding.} 
     To capture contextual dependencies within each modality, we apply independent Transformer encoders as follows. Each encoder uses self-attention, where the query, key, and value vectors are all derived from the same modality.
     \begin{equation}
         \mathbf{h}_{\text{Ma}}^{(m)} = \text{Attention}^{(m)}(\mathbf{h}^{(m)}, \mathbf{h}^{(m)}, \mathbf{h}^{(m)}),
     \end{equation}
     
     where \( \mathbf{h}_{\text{Ma}}^{(m)} \in \mathbb{R}^{d} \) encodes intra-modal semantic.

    \textbf{Interaction-aware Encoding.}
    To facilitate efficient cross-modal interaction, we introduce a gated cross-attention mechanism. 
    For each query modality \(m\), we take its modality‑specific encoded representation 
    \(\mathbf{h}^{(m)} \in \mathbb{R}^{d}\) from the previous stage as the query vector 
    \(\mathbf{q}^{(m)} = \mathbf{h}^{(m)}\). 
    The remaining modalities are indexed by \(\tilde{m} \in\{\text{text},\text{audio},\text{visual}\}-\{m\} \).

    We first compute gating vectors to modulate the influence of non‑query modalities.
    \begin{equation}
    \mathbf{g}^{(\tilde{m})}
       = \sigma\ \!\bigl(\mathbf{W}^{(\tilde{m})}\mathbf{q}^{(m)} + \mathbf{b}^{(\tilde{m})}\bigr),
    \end{equation}

    where $\mathbf{W}^{(\tilde{m})} \in \mathbb{R}^{d \times d}$ and 
    $\mathbf{b}^{(\tilde{m})} \in \mathbb{R}^{d}$ are learnable parameters, 
    and $\sigma(\cdot)$ denotes the sigmoid activation.
    The gated representations are aggregated by an element‑wise (Hadamard) product,
    \begin{equation}
    \mathbf{k}^{(m)} 
       = \sum_{\tilde{m}}
         \mathbf{g}^{(\tilde{m})} \odot \mathbf{h}^{(\tilde{m})}, 
    \quad
    \mathbf{v}^{(m)} = \mathbf{k}^{(m)}.
    \end{equation}

    The cross-modal context is then integrated through scaled dot-product attention.
    \begin{equation}
    \mathbf{h}_{\text{Ia}}^{(m)} = \text{Attention}^{(m)}(\mathbf{q}^{(m)}, \mathbf{k}^{(m)}, \mathbf{v}^{(m)}),
    \end{equation}
    yielding interaction-aware features \( \mathbf{h}_{\text{Ia}}^{(m)} \in \mathbb{R}^{d} \) that capture selectively modulated dependencies across modalities.
    
    \textbf{Contextual Feature Integration.}
    Finally, we combine the outputs of both encoding stages. The contextual vector \( \mathbf{h}_{\text{Ma}}^{(m)} \) and the interaction-aware vector \( \mathbf{h}_{\text{Ia}}^{(m)} \) are concatenated and projected into the unified representation space as
    \begin{equation}
    \tilde{\mathbf{h}}^{(m)} = \mathbf{W}^{(m)} \begin{bmatrix} \mathbf{h}_{\text{Ma}}^{(m)} 
    \\ \mathbf{h}_{\text{Ia}}^{(m)} \end{bmatrix} + \mathbf{b}^{(m)},
    \end{equation}

    where \( \mathbf{W}^{(m)} \in \mathbb{R}^{d \times 2d} \) and \( \mathbf{b}^{(m)} \in \mathbb{R}^{d} \) are modality-specific parameters. The resulting vector \( \tilde{\mathbf{h}}^{(m)} \in \mathbb{R}^{d} \) integrates both intra- and cross-modal information, and is subsequently passed to the unimodal fusion module for further processing.

    \subsection{Representation-Space Fusion}
   
    To model high-order interactions among different modalities, we propose the \textbf{Synergistic Polynomial Fusion (SPF)} module, inspired by polynomial tensor pooling~\cite{NEURIPS2019_f56d8183} and canonical polyadic (CP) decomposition. SPF simulates multi‑way multiplicative interactions through low‑rank projections, enhancing expressive power without explicitly constructing high‑dimensional tensors. Different from prior works, we replace the shared projections with \emph{modality-specific} ones, augment them with static gating, and apply a stabilizing non-linearity. enabling asymmetric, noise‑aware high‑order interactions while remaining compact. The input to SPF is the modality‑specific contextual representation $\tilde{\mathbf{h}}^{(m)} \in \mathbb{R}^{d}$ obtained from the preceding two‑stage encoding process.

    To preserve modality-specific characteristics and simulate polynomial interactions of order \( p \), we apply a set of independent linear projections to each modality. For interaction order \( j \in \{1, \dots, p\} \), each modality \( m \) is projected as follows.
    \begin{equation}
    \mathbf{z}_j^{(m)} = \phi\left( \mathbf{W}_j^{(m)} \tilde{\mathbf{h}}^{(m)} \right), \quad \mathbf{W}_j^{(m)} \in \mathbb{R}^{r \times d},
    \end{equation}
    
    where \( \phi(\cdot) \) denotes the GELU activation and \( r \) is the projection rank. To regulate the contribution of each modality, we apply a static gating mechanism with learnable scalar weights \( \lambda^{(m)}_{j} \in (0, 1) \). We also include a bias vector \( \mathbf{w}_j \in \mathbb{R}^{r} \), which simulates the constant path in CP factorization and is added directly rather than concatenated. This constant path acts as a stabilizing anchor, compensating for the multiplicative amplification introduced by high-order interactions and preserving baseline semantics even when modality-specific signals are weak or noisy.
    
    The gated fusion vector for each order is computed as
    \begin{equation}
    \mathbf{z}_j = \sum_{m} \lambda^{(m)}_{j} \mathbf{z}^{(m)}_{j} + \mathbf{w}_j,
    \end{equation}
    
    where \( \mathbf{z}_j \in \mathbb{R}^{r} \). These vectors are then aggregated via element-wise multiplication, i.e., \( \mathbf{z} = \mathbf{z}_1 \odot \mathbf{z}_2 \odot \cdots \odot \mathbf{z}_p \), where \( \mathbf{z} \in \mathbb{R}^{r} \) denotes a rank-1 approximation of an implicit \( p \)-order multimodal tensor. To stabilize the fused representation, a signed square root transformation is applied.
    \begin{equation}
    \hat{\mathbf{z}} = \text{sign}(\mathbf{z}) \odot \sqrt{|\mathbf{z}|}.
    \end{equation}
    
    where $|\cdot|$ and $\sqrt{\cdot}$ are applied element‑wise.
    
    This transformation controls the dynamic range of the fused tensor by reducing magnitude variance while retaining directional cues, which facilitates more stable gradient propagation during backpropagation.
    
    Finally, the normalized vector \( \hat{\mathbf{z}} \in \mathbb{R}^{r} \) is projected back to the shared representation space and passed to a softmax classifier.  
    \begin{equation}
    \hat{\mathbf{y}}
    = \operatorname{softmax}\!\left(
    \mathbf{W}_{\text{cls}}^{\top}\big(\mathbf{W}_{\text{out}}^{\top}\hat{\mathbf{z}}+\mathbf{b}_{\text{out}}\big)
    + \mathbf{b}_{\text{cls}}
    \right),
    \end{equation}
    
    where $\hat{\mathbf{z}}\in\mathbb{R}^{r}$, 
    $\mathbf{W}_{\text{out}}\in\mathbb{R}^{r\times d}$, 
    $\mathbf{b}_{\text{out}}\in\mathbb{R}^{d}$, 
    $\mathbf{W}_{\text{cls}}\in\mathbb{R}^{d\times c}$, 
    $\mathbf{b}_{\text{cls}}\in\mathbb{R}^{c}$,
    with $c$ denoting the number of emotion classes and $d$ the output dimension of the fusion layer.

    The multimodal classification loss is computed with a mask-aware cross‑entropy.
    
    \begin{equation}
\mathcal{L}_{\text{1}}
= -\frac{1}{\sum_{i}\mu_i}\;
  \sum_{i=1}^{N}\mu_i
  \sum_{j=1}^{c} y_{i,j}\,\log\bigl(\hat{y}_{i,j}\bigr),
\end{equation}

    where \(N\) is the total number of utterances, \(c\) the number of emotion classes, 
\(y_{i,j}\in\{0,1\}\) is a one-hot ground-truth indicator that equals 1 when utterance \(i\) belongs to class \(j\), 
\(\hat{y}_{i,j}\) is the predicted probability of class \(j\) for utterance \(i\), 
and \(\mu_i\in\{0,1\}\) is a binary mask that sets padding or otherwise invalid samples to 0.

    \subsection{Gradient-Space Optimization}
    We frame training as a multi‑objective problem that combines the primary classification loss with auxiliary supervision. To balance these objectives, we introduce the \textbf{Pareto Gradient Modulator (PGM)}, which selects descent directions that are Pareto‑optimal, thereby reducing gradient interference and improving training stability. Unlike static or manually tuned weights, PGM dynamically adapts the relative importance of each objective according to the current gradient geometry. This dynamic re‑weighting enables principled trade‑offs among tasks and effectively mitigates optimization conflicts.

\textbf{Unimodal Regularization.}
 Following the self-distillation framework introduced in~\cite{ma2024sdt}, we regularize each unimodal branch by using the fused prediction as a soft teacher. Each branch receives both hard-label supervision and soft-label alignment. The unimodal classification loss is defined as
\begin{equation}
\mathcal{L}_{2}
= -\frac{1}{\sum_{i}\mu_i}
  \sum_{i=1}^{N}\mu_i
  \sum_{m}
  \sum_{j=1}^{c}
    y_{i,j}\,\log\bigl(\hat{y}^{(m)}_{i,j}\bigr),
\end{equation}

where \(\mu_i\in\{0,1\}\) masks out padded or otherwise invalid utterances,  
\(y_{i,j}\) is the one-hot ground-truth indicator for class \(j\), and  
\(\hat{y}^{(m)}_{i,j}\) is the probability that modality \(m\) assigns to that class.

The distillation loss is computed as
\begin{equation}
\mathcal{L}_{3}
  = \frac{1}{\sum_{i}\mu_i}
    \sum_{i=1}^{N}\mu_i
    \sum_{m}
      \mathrm{KL}\!\bigl(\tilde{\mathbf{y}}^{(\mathrm{all})}_i
                         \,\big\|\, 
                         \tilde{\mathbf{y}}^{(m)}_i\bigr),
\end{equation}

where $\tilde{\mathbf{y}}=\mathrm{softmax}(\hat{\mathbf{y}}/T)$ denotes the temperature-scaled ($T$) probability vector; $\tilde{\mathbf{y}}^{(\mathrm{all})}_i$ is the teacher distribution from the fused branch, and $\tilde{\mathbf{y}}^{(m)}_i$ is the student distribution produced by modality~$m$.  
The divergence is computed as $\mathrm{KL}(P\!\parallel\!Q)=\sum_{k=1}^{c} P_k \log\!\bigl(P_k/Q_k\bigr)$, where $k$ indexes the $c$ emotion classes.

    \textbf{Dynamic Gradient Coordination.}
    Inspired by the multi-objective perspective of gradient conflicts introduced by MGDA \cite{sener2018multi}, we treat the three supervision signals in MERC—multimodal classification
    \((\mathcal{L}_{1})\), unimodal regularization \((\mathcal{L}_{2})\), and distillation
    \((\mathcal{L}_{3})\)—as competing objectives.
    To harmonize them, we introduce the \textbf{Pareto Gradient Modulator (PGM)}. Unlike MGDA, which must iteratively solve a quadratic program at every update, PGM normalizes the task gradients and then optimizes a fixed three‑variable QP. Its low dimensionality keeps the overhead small while producing a Pareto‑descent direction that balances the three objectives, effectively mitigating gradient conflicts and stabilizing training.

    To compute task-wise trade-offs, we first calculate the gradients \( \{ \nabla_\theta \mathcal{L}_{\text{1}}, \nabla_\theta \mathcal{L}_{\text{2}}, \nabla_\theta \mathcal{L}_{\text{3}} \} \) with respect to the shared model parameters \( \theta \). Each \( \nabla_\theta \mathcal{L}_i \in \mathbb{R}^d \) is a vector in the shared parameter space of dimension \( d \), and Pareto optimality is defined over this multi-dimensional gradient space.

    Let \( \boldsymbol{\gamma} = [\gamma_{\text{1}}, \gamma_{\text{2}}, \gamma_{\text{3}}] \in \mathbb{R}^3 \) denote the task weights used to balance the three objectives. We seek a Pareto-optimal descent direction by minimizing the squared norm of the weighted gradient sum.
    \begin{equation}
    \min_{\boldsymbol{\gamma}} \left\| \sum_{i=1}^3 \gamma_i \nabla_\theta \mathcal{L}_i \right\|_2^2, \quad \; s.t. \; \boldsymbol{\gamma} \succeq \mathbf{0},\; \mathbf{e}_3^\top \boldsymbol{\gamma} = 1,
    \end{equation}
    
    where \( \mathbf{e}_3 = [1, 1, 1]^\top \) denotes an all-one column vector.
    
    This objective can be equivalently formulated as a quadratic programming (QP) problem:
    \begin{equation}
    \min_{\boldsymbol{\gamma}} \; \frac{1}{2} \boldsymbol{\gamma}^\top Q \boldsymbol{\gamma}, \qquad 
    s.t. \; \boldsymbol{\gamma} \succeq \mathbf{0},\; \mathbf{e}_3^\top \boldsymbol{\gamma} = 1,
    \end{equation}

    where \(\mathbf{g}_i\in\mathbb{R}^{d}\) denotes the \(L_2\)-normalized gradient of task \(i\)
 with respect to the shared parameters \(\theta\).
Stacking them yields 
\(G=[\,\mathbf{g}_1,\mathbf{g}_2,\mathbf{g}_3]^{\top}\in\mathbb{R}^{3\times d}\),
and the task‑wise gradient inner‑product matrix is 
\(Q = GG^{\top}\in\mathbb{R}^{3\times3}\).

    In practice, we construct \(G\) from the normalized gradients at each mini‑batch
    and solve the QP with a lightweight sequential quadratic‑programming routine to
    dynamically adjust the task weights.
    
    The resulting task weights \( \boldsymbol{\gamma} \) are then used to compute the composite training loss:
    \begin{equation}
    \mathcal{L} = \gamma_{\text{1}} \cdot \mathcal{L}_{\text{1}} + \gamma_{\text{2}} \cdot \mathcal{L}_{\text{2}} + \gamma_{\text{3}} \cdot \mathcal{L}_{\text{3}}.
    \end{equation}
    
    By coordinating learning signals at the gradient level, PGM complements SPF’s representation-level fusion, jointly contributing to the overall stability and effectiveness of the Cross-Space Synergy framework.

        \begin{table*}[t]
  \centering

  \begin{minipage}{\textwidth}
    \centering
    \fontsize{8.5pt}{12pt}\selectfont
    \setlength{\tabcolsep}{6pt}
    \begin{tabular}{lcccccccccccccc}
    \toprule
    \multirow{2}{*}{\textbf{Models}} & \multicolumn{2}{c}{happy} & \multicolumn{2}{c}{sad} & \multicolumn{2}{c}{neutral} & \multicolumn{2}{c}{angry} & \multicolumn{2}{c}{excited} & \multicolumn{2}{c}{frustrated} & \multirow{2}{*}{\textbf{ACC}} & \multirow{2}{*}{\textbf{w-F1}} \\
     & ACC & F1 & ACC & F1 & ACC & F1 & ACC & F1 & ACC & F1 & ACC & F1 \\
    \midrule
        CMN         & 24.31 & 30.30 & 56.33 & 62.02 & 52.34 & 52.41 & 61.76 & 60.17 & 56.19 & 60.76 & \underline{\textbf{72.44}} & 61.27 & 56.87 & 56.33 \\
        ICON        & 25.00 & 31.30 & 67.35 & 73.17 & 55.99 & 58.50 & 69.41 & 66.29 & 70.90 & 67.09 & 71.92 & 65.08 & 62.85 & 62.25 \\
        DialogueRNN & 25.00 & 34.95 & 82.86 & 84.58 & 54.43 & 57.66 & 61.76 & 64.42 & \underline{\textbf{90.97}}\ & 76.30 & 62.20 & 59.55 & 65.43 & 64.29 \\
        MMGCN       & 32.64 & 39.66 & 72.65 & 76.89 & 65.10 & 62.81 & 73.53 & \underline{\textbf{71.43}}\ & 77.93 & 75.40 & 65.09 & 63.43 & 66.61 & 66.25 \\
        MM-DFN & 44.44 & 44.44 & 77.55 & 80.00 & 71.35 & 66.99 & \underline{\textbf{75.88}} & 70.88 & 74.25 & 76.42 & 58.27 & 61.67 & 67.84 & 67.85 \\
        MPT-HCL      & - & 58.13 & - & \underline{\textbf{85.97}} & - & 66.75 & - & 69.97 & - & 74.06 & - & 69.06 & 72.83 & 72.51 \\
        M3NET     & 58.78 & 56.00 & 76.23 & 76.07 & 68.75 &68.75 & 65.22 & 67.80 & 78.93 & 78.93 & 64.57 & 64.57 & 69.56 & 69.52 \\
        GraphSmile  & \underline{\textbf{62.88}}     & 60.14 & 80.72     & 81.38 & 62.63     & 68.48 & 71.52     & 70.45 & 82.09     & 77.60 & 69.36     & 66.02 & 70.98     & 71.00 \\
        HAUCL     & 48.10     & 50.33     & 83.05    & 81.50     & 65.68     & 67.43     & 74.05     & 64.45     & 77.24     & 73.02     & 63.53     & 67.00     & 68.52 & 68.65 \\
        Ada2I        & 48.51     & 46.76     & 81.93     & 80.75     & 68.33     & 64.28     & 69.14     & 67.47     & 70.09     & 73.65     & 64.27     & 67.17     & 68.08 & 67.95 \\
        CFN-ESA      & 50.56     & 56.17     & 79.53     &80.96     & 64.76     & 68.94     & 70.70     & 67.89     & 81.78     & 73.99     & 68.68     & 65.57     & 69.50 & 69.65 \\
        SDT              & 58.19 & 64.17 & 85.59 & 81.37 & 75.34 & 74.24 & 65.61 & 69.08 & 84.56 & 82.53 & 70.03 & 69.66 & 74.12 & 74.34 \\
    \midrule
        \textbf{Ours}    & 58.43 & \underline{\textbf{64.60}} &\underline{\textbf{86.28}}  & 82.20 &\underline{\textbf{77.60}}  & \underline{\textbf{76.68}} & 69.64 & 69.23 & 87.08 & \underline{\textbf{82.81}} & 69.38 & \underline{\textbf{71.50} } & \underline{\textbf{75.42}} & \underline{\textbf{75.66}} \\
    \bottomrule
    \end{tabular}
    \caption{Performance comparison on the IEMOCAP dataset.  
    Underlined numbers indicate the best performance and “–” indicates unreported metrics in the original work.}
        \label{tab:iemocap_results}
  \end{minipage}

  \vspace{1em}

  \begin{minipage}{\textwidth}
    \centering
    \fontsize{8.5pt}{12pt}\selectfont
    \setlength{\tabcolsep}{4.15pt}
    \begin{tabular}{lcccccccccccccccc}
    \toprule
    \multirow{2}{*}{\textbf{Models}} & \multicolumn{2}{c}{anger} & \multicolumn{2}{c}{disgust} & \multicolumn{2}{c}{fear} & \multicolumn{2}{c}{joy} & \multicolumn{2}{c}{neutral} & \multicolumn{2}{c}{sadness} & \multicolumn{2}{c}{surprise} & \multirow{2}{*}{\textbf{ACC}} & \multirow{2}{*}{\textbf{w-F1}} \\
     & ACC & F1 & ACC & F1 & ACC & F1 & ACC & F1 & ACC & F1 & ACC & F1 & ACC & F1 \\
    \midrule
        DialogueRNN           & \underline{\textbf{82.17}} & 76.56 & 46.62 & 47.64 & 0.00 & 0.00 & 21.15 & 24.65 & 49.50 & 51.49 & 0.00 & 0.00 & 48.41 & 46.01 & 60.27 & 57.95 \\
        MMGCN           & 84.32 & 76.96 & 47.33 & 49.63 & 2.00 & 3.64 & 14.90 & 20.39 & 56.97 & 53.76 & 1.47 & 2.82 & 42.61 & 45.23 & 61.34 & 58.41 \\
        MM-DFN           & 79.06 & 75.80 & 53.02 & 50.42 & 0.00 & 0.00 & 17.79 & 23.72 & 59.20 & 55.48 & 0.00 & 0.00 & 50.43 & 48.27 & 60.96 & 58.72 \\
        MPT-HCL           & - & 77.82 & - & 58.26 & - & 21.52 & - & 45.15 & - & 60.18 & - & 30.36 & - & 59.25 & 65.86 & 65.02 \\
        M3NET           & 75.07 & 79.99 & \underline{\textbf{61.09}} & \underline{\textbf{60.43}} & 30.77 & 21.05 & 50.88 & 36.02 & 65.36 & 63.87 & 32.00 & 27.12 & 55.02 & 53.71 & 67.32 & 65.91 \\
        GraphSmile           & 75.77 & 79.95 & 52.94 & 56.62 & \underline{\textbf{32.14}} & \underline{\textbf{23.08}} & 57.41 & 39.24 &  \underline{\textbf{68.62}} & 62.99 & 41.67 & 28.85 & 50.13 & 51.96 & 66.67 & 65.46 \\
        HAUCL            & 75.46 & 79.93 & 59.01 & 59.22 & 24.24 & 19.28 & 56.60 & 38.22 & 66.49 & 64.00 & 37.84 & 26.67 & 52.47 & 53.88 & 67.24 & 65.93 \\
        Ada2I            & 71.31 & 77.83 & 55.00 & 50.67 & 0.00 & 0.00 & 47.78 & 28.86 & 54.57 & 54.77 & 0.00 & 0.00 & 47.81 & 49.23 & 63.10 & 60.15  \\
        CFN-ESA           & 75.92 & 79.77 & 57.34 & 58.54 & 30.43 & 19.18 & 56.76 & 39.50 & 64.76 & \underline{\textbf{64.84}} & 38.84 & 28.04 & 54.47 & 54.62 & 67.43 & 66.15 \\
        SDT           & 76.73 & 80.20 & 56.72  & 59.04 & 29.41 & 14.93 & 54.35 & 43.35 & 64.69 & 63.54 & 39.29 & 22.92 & 53.37 & 53.41 & 67.36 & 66.14\\
    \midrule
        \textbf{Ours}  & 77.20 & \underline{\textbf{80.74}} & 56.86 & 58.62 & 28.00 & 18.67 & \underline{\textbf{62.10}} & \underline{\textbf{46.39}} & 65.24 & 64.83 & \underline{\textbf{44.44}} & \underline{\textbf{30.77}} & \underline{\textbf{55.40}} & \underline{\textbf{55.95}} & \underline{\textbf{68.47}} & \underline{\textbf{67.41}} \\
    \bottomrule
    \end{tabular}
    \caption{Performance comparison on the MELD dataset. }
     \label{tab:meld_results}
  \end{minipage}

\end{table*}   
    \section{Experiments}
    \subsection{Experimental Setup}
    We evaluate our model on two widely-used benchmarks for MERC: \textbf{IEMOCAP}~\cite{2008IEMOCAP} and \textbf{MELD}~\cite{poria-etal-2019-meld}. IEMOCAP contains approximately 12 hours of dyadic conversations annotated with six emotion categories, along with aligned audio and visual modalities. MELD, derived from the TV series \textit{Friends}, consists of multi-party conversations labeled at the utterance level with seven emotion classes. We follow the standard train/validation/test splits and classification settings for both datasets.
    
    For fair comparison, we use utterance-level features consistent with prior work~\cite{ghosal-etal-2020-cosmic, hu-etal-2021-mmgcn, ma2024sdt}. Text features (\( \mathbf{x}_t^{(t)} \in \mathbb{R}^{1024} \)) are obtained as [CLS] embeddings from a fine-tuned RoBERTa-Large model. Audio features (\( \mathbf{x}_t^{(a)} \in \mathbb{R}^{1582} \) for IEMOCAP and \( \mathbb{R}^{300} \) for MELD) are extracted using openSMILE (IS10). Visual features (\( \mathbf{x}_t^{(v)} \in \mathbb{R}^{342} \)) are derived from a DenseNet pretrained on FER+. The model is implemented in PyTorch and trained using the Adam optimizer with a batch size of 32 on a single RTX 2080Ti GPU. For IEMOCAP, we set the learning rate to \( 1\mathrm{e}{-4} \) and the distillation temperature to \( T{=}1 \); for MELD, \( 5\mathrm{e}{-6} \) and \( T{=}4 \). The Transformer encoder contains one layer with 8 heads and 1024 hidden/output dimensions. Dropout is set to 0.5, weight decay to \( 1\mathrm{e}{-5} \), and the fusion module adopts order \( p{=}3 \) for IEMOCAP and \( p{=}4 \) for MELD, with rank \( r{=}16 \) and output dimension 1024.

    \begin{figure*}[t]
        \centering
        \includegraphics[width=\textwidth]{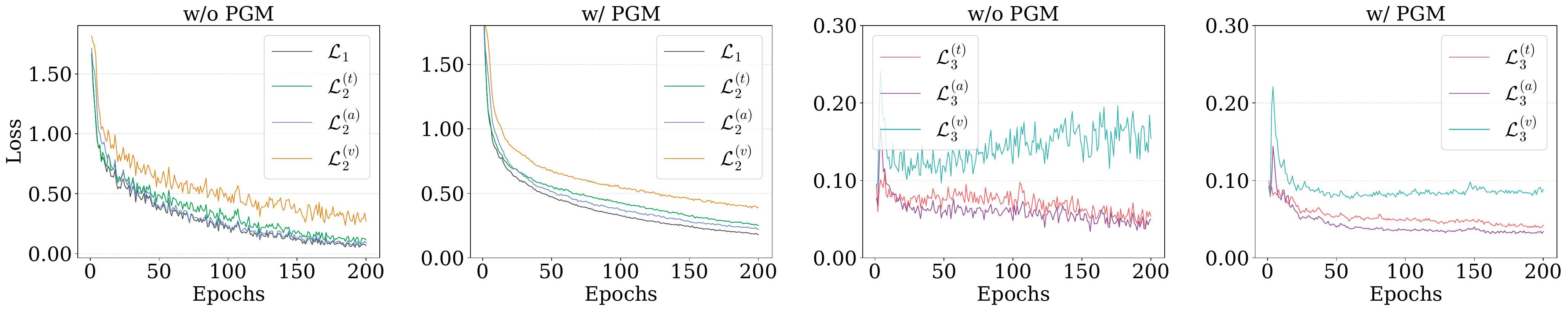}
        \caption{
        Training loss curves on IEMOCAP with and without PGM. The left two plots show multimodal loss \( \mathcal{L}_1 \) and unimodal losses \( \mathcal{L}_2^{(m)} \); the right show KL losses \( \mathcal{L}_3^{(m)} \). PGM improves convergence and reduces variance, especially for KL losses.
        }
        \label{fig:losses}
    \end{figure*}
    
    \subsection{Baseline Details}
    We compare CSS with strong baselines across three mainstream paradigms in MERC. (1) \textbf{Sequence-based models}  ICON~\cite{hazarika-etal-2018-icon}, CMN~\cite{hazarika-etal-2018-conversational}, DialogueRNN~\cite{Majumder2018DialogueRNNAA} use RNNs with speaker-aware encoding and early or unimodal fusion. (2) \textbf{Graph-based models} MMGCN~\cite{hu-etal-2021-mmgcn}, MM-DFN~\cite{hu2022mm}, M3NET~\cite{10203083}, GraphSmile~\cite{li2025tracing}, HAUCL~\cite{yi2024multimodal} apply GNNs to capture structural dependencies but rely on fixed message-passing schemes. (3) \textbf{Attention-based models} MPT-HCL~\cite{zou2023multimodal}, Ada2I~\cite{nguyen2024ada2i}, CFN-ESA~\cite{10502283}, SDT~\cite{ma2024sdt} focus on cross-modal attention but often face modality entanglement or unstable training.
    
    We adopt standard MERC evaluation metrics~\cite{ghosal-etal-2019-dialoguegcn,Majumder2018DialogueRNNAA}, including overall accuracy (ACC), weighted F1-score (w-F1), and per-class scores. Tables~\ref{tab:iemocap_results} and~\ref{tab:meld_results} present comparative results. For recent open-source methods, we reproduce results using their official configurations; for others, we report values from original papers or reliable reproductions.

    \subsection{Interpretation and Discussion}
    Results on MERC are summarised in Tables~\ref{tab:iemocap_results} and~\ref{tab:meld_results}{}. 
    Our \textbf{Cross‑Space Synergy (CSS)} attains the best overall performance on both \textbf{IEMOCAP} and \textbf{MELD} without the severe per‑class fluctuations reported in earlier work 
    (e.g., CMN’s markedly high \emph{frustrated} but low \emph{happy}, or DialogueRNN’s bias toward \emph{excited}). 
    The comparison also highlights the limited ability of early sequence-based models to capture complex multimodal interactions, explaining why they now lag behind newer attention- and graph-based architectures. 
    Across all emotions, CSS either matches or surpasses the strongest existing scores, and no category shows a noticeable shortfall, underscoring the balanced nature of our approach.

    In the six-class IEMOCAP test, CSS achieves the highest ACC (\textbf{75.42\%}) and w-F1 (\textbf{75.66\%}).
It takes the lead in at least one metric among the five emotions—\emph{happy}, \emph{sad}, \emph{neutral}, \emph{excited}, and \emph{frustrated}.
For \emph{neutral}, CSS surpasses the SDT baseline by 2.26 \% in ACC and 2.44 \% in w-F1, which we attribute to the expressiveness of SPF and the adaptive weighting of PGM that preserve subtle contextual cues.
Graph-based models generally have a slight advantage for \emph{angry}, suggesting their structural bias favours this emotion. In contrast, attention-based approaches are less advantageous for this class. Even so, CSS maintains a close performance with the lead, indicating that its representations are broad rather than over-specialised.
    
    The seven‑class MELD benchmark is larger and more imbalanced, yet CSS also achieves the best overall ACC (\textbf{68.47\%}) and w‑F1 (\textbf{67.41\%}).  
    It achieves at least one leading metric for \emph{anger}, \emph{joy}, \emph{sadness}, and \emph{surprise}; on \emph{joy}, it improves ACC by \(4.69\%\) and F1 by \(3.04\%\) over the previous best.  
    All methods—including ours—struggle on the extremely minority \emph{fear} and \emph{disgust} classes (each accounts for about 2\% of the training data), but CSS still posts the best or near‑best values.  
    Handling this scarcity will be left for future work via data augmentation or uncertainty weighting.
    
    Training efficiency was also measured under the same input settings.  
    Per‑epoch times (CSS 2.36s, SDT 1.70s, CFN‑ESA 5.42s, GraphSmile 7.10s, M3NET 42.75s) show that CSS is only marginally slower than SDT; the additional 0.66s stems from the small QP solved by PGM, yet yields sizeable performance gains.  
    Compared to other attention-based or graph-based models, CSS is both faster and more accurate.  
    Importantly, PGM is used only during training, so inference speed is unaffected.
    
    In summary, CSS raises the performance ceiling while safeguarding the floor: it never sacrifices one emotion for another, and the gap between majority and minority classes narrows.  
    At the same time, it maintains competitive efficiency.  
    These outcomes confirm that coupling expressive, high-order fusion in representation space with gradient-space coordination is crucial for achieving balanced and robust multimodal emotion recognition.

    \begin{table}[t]
    \centering
    \fontsize{8.5pt}{12pt}\selectfont
    \setlength{\tabcolsep}{8pt}
    \begin{tabular}{lcccc}
        \toprule
        \multirow{2}{*}{\textbf{Model Variant}} & \multicolumn{2}{c}{\textbf{IEMOCAP}} & \multicolumn{2}{c}{\textbf{MELD}} \\
        \cmidrule(r){2-3} \cmidrule(r){4-5}
        & ACC & w-F1 & ACC & w-F1 \\
        \midrule
        \textbf{CSS (Ours)} & \textbf{75.42} & \textbf{75.66} & \textbf{68.47} & \textbf{67.41} \\
        \midrule
        w/o SPF & 74.86 & 75.15 & 67.85 & 66.79 \\
        w/o MSP$^\dagger$ & 74.37 & 74.46 & 67.13 & 65.79 \\
        w/o PGM & 74.61 & 74.66 & 67.20 & 65.74 \\
        w/o MAE & 73.57 & 73.54 & 66.97 & 65.95 \\
        w/o IAE & 70.98 & 71.07 & 67.01 & 65.66 \\
        w/o $\mathcal{L}_{\text{2}}$ & 74.18 & 74.30 & 67.93 & 66.79 \\
        w/o $\mathcal{L}_{\text{3}}$ & 72.64 & 72.76 & 66.82 & 66.70 \\
        \bottomrule
    \end{tabular}
    \caption{Ablation study on IEMOCAP and MELD. MSP = Modality-Specific Projection; MAE/IAE = Modality-/Interaction-aware Encoding. †Static gating is removed when MSP is ablated.
    }
    
    \label{tab:ablation_combined}
        
    \centering
 
    \end{table}

    \begin{table}[t] 
    \centering
    \fontsize{8.5pt}{12pt}\selectfont
    \setlength{\tabcolsep}{1pt}
    \begin{tabular}{
        >{\centering\arraybackslash}m{1cm} 
        >{\centering\arraybackslash}m{1cm} 
        >{\centering\arraybackslash}m{1cm} 
        >{\centering\arraybackslash}m{1.1cm} 
        >{\centering\arraybackslash}m{1.1cm} 
        >{\centering\arraybackslash}m{1.1cm} 
        >{\centering\arraybackslash}m{1.1cm}
    }
        \toprule
        \multicolumn{3}{c}{\textbf{Modality Setting}} & 
        \multicolumn{2}{c}{\textbf{IEMOCAP}} & 
        \multicolumn{2}{c}{\textbf{MELD}} \\
        \cmidrule(lr){1-3} \cmidrule(lr){4-5} \cmidrule(lr){6-7}
        \textbf{Text} & \textbf{Audio} & \textbf{Visual} & ACC & w-F1 & ACC & w-F1 \\
        \midrule
        $\checkmark$ & $\checkmark$ & $\checkmark$ & \textbf{75.42} & \textbf{75.66} & \textbf{68.47} & \textbf{67.41} \\
        \midrule
        $\times$ & $\checkmark$ & $\checkmark$ & 61.43 & 61.34 & 49.46 & 39.13 \\
        $\checkmark$ & $\times$ & $\checkmark$ & 68.21 & 68.59 & 66.82 & 65.77 \\
        $\checkmark$ & $\checkmark$ & $\times$ & 68.21 & 68.57 & 66.67 & 65.51 \\
        $\checkmark$ & $\times$ & $\times$ & 66.05 & 66.09 & 66.28 & 65.36 \\
        $\times$ & $\checkmark$ & $\times$ & 61.74 & 62.39 & 49.00 & 36.50 \\
        $\times$ & $\times$ & $\checkmark$ & 42.64 & 43.21 & 48.12 & 31.28 \\
        \bottomrule
    \end{tabular}
    \caption{Modality Ablation Results.}
    \label{modality_ablation}
    
    \end{table}

    \subsection{Ablation Studies}
   To gauge the contribution of each CSS component, we selectively remove or replace it, and report the results in Table \ref{tab:ablation_combined}.  
    Disabling any single module causes a noticeable drop in both ACC and w‑F1 on IEMOCAP and MELD, confirming that the components are complementary rather than redundant.  
  Removing the attention structure—interaction‑aware encoding (IAE) and its modality‑aware encoding (MAE)—both lead to a significant decline, underscoring the importance of the attention mechanism for MERC.  
    Eliminating Synergistic Polynomial Fusion (SPF) or its modality‑specific projection (MSP) further degrades results, highlighting the value of high‑order, asymmetric fusion.  
    Removing the Pareto Gradient Modulator (PGM) also lowers overall scores, indicating that adaptive gradient coordination is critical for stable optimization.  
    Finally, dropping the auxiliary objectives $\mathcal{L}_{\text{2}}$ and $\mathcal{L}_{\text{3}}$ harms generalisation, showing that regularisation and distillation guide SPF to discriminative representations.

    Figure~\ref{fig:losses} contrasts training dynamics with and without PGM.  
    When PGM is enabled, both multimodal and unimodal loss curves become noticeably smoother. Especially $L_{3}^{(v)}$  showed a convergence trend under PGM optimization. PGM suppresses the oscillations that arise with fixed task weights and thereby enhances overall optimization stability.

    Finally, we examine the impact of withholding individual modalities, and the corresponding results are presented in Table \ref{modality_ablation}.  Removing any single modality degrades
    overall performance, and retaining only one modality causes a pronounced
    collapse.  These results show
    that CSS genuinely leverages complementary cues across text, audio, and visual
    streams rather than over‑fitting to a dominant modality, validating its design
    for balanced multimodal fusion.

    \section{Conclusion}
    This paper proposes a unified MERC framework CSS that combines \textit{Synergistic Polynomial Fusion}—a stabilized low‑rank, high‑order fusion—with \textit{Pareto Gradient Modulator}, which follows locally Pareto‑optimal directions to curb gradient conflicts. On IEMOCAP and MELD, CSS yields consistent gains with lower variance and competitive cost. Future work will explore broader multimodal settings and pursue a more expressive fusion structure with stronger conflict‑resilient optimization.

\bibliography{aaai2026}

\end{document}